\documentclass[conference]{IEEEtran}
\IEEEoverridecommandlockouts
\usepackage{cite}
\usepackage{amsmath,amssymb,amsfonts}
\usepackage{algorithmic}
\usepackage{graphicx}
\usepackage{textcomp}
\usepackage{mathtools}

\DeclarePairedDelimiter\avg{E[}{]}

\def\BibTeX{{\rm B\kern-.05em{\sc i\kern-.025em b}\kern-.08em
    T\kern-.1667em\lower.7ex\hbox{E}\kern-.125emX}}

\usepackage{tikz}

\newcommand\copyrighttext{%
	\footnotesize \textcopyright 2018 IEEE. Personal use of this material is permitted.
	Permission from IEEE must be obtained for all other uses, in any current or future
	media, including reprinting/republishing this material for advertising or promotional
	purposes, creating new collective works, for resale or redistribution to servers or
	lists, or reuse of any copyrighted component of this work in other works.}
\newcommand\copyrightnotice{%
	\begin{tikzpicture}[remember picture,overlay]
	\node[anchor=south,yshift=10pt] at (current page.south) {\fbox{\parbox{\dimexpr\textwidth-\fboxsep-\fboxrule\relax}{\copyrighttext}}};
	\end{tikzpicture}%
}

\begin{document}

\title{Analytical Study of Incremental Approach for Information Dissemination in Wireless Networks\thanks{The research was done at IITP RAS and supported by the Russian Science Foundation (agreement No 14-50-00150).}
}

\author{\IEEEauthorblockN{Andrey Belogaev\IEEEauthorrefmark{1}, Evgeny Khorov\IEEEauthorrefmark{1}\IEEEauthorrefmark{2}, Artem Krasilov\IEEEauthorrefmark{1}\IEEEauthorrefmark{2}, Andrey Lyakhov\IEEEauthorrefmark{1}}
\IEEEauthorblockA{\IEEEauthorrefmark{1}Institute for Information Transmission Problems, Russian Academy of Sciences, Moscow, Russia\\ 
\IEEEauthorrefmark{2}National Research University Higher School of Economics, Moscow, Russia\\
E-mail: \{belogaev, khorov, krasilov, lyakhov\}@iitp.ru}
}

\maketitle

\copyrightnotice

\begin{abstract}
In many scenarios, control information dissemination becomes a bottleneck, which limits the scalability and the performance of wireless networks. 
Such a problem is especially crucial in mobile ad hoc networks, dense networks, networks of vehicles and drones, sensor networks. In other words, this problem occurs in any scenario with frequent changes in topology or interference level on one side and with strong requirements on delay, reliability, power consumption, or capacity on the other side. 
If the control information changes partially, it may be worth sending only differential updates instead of messages containing full information to reduce overhead. However, such an approach needs accurate tuning of dissemination parameters, since it is necessary to guarantee information relevance in error-prone wireless networks. In the paper, we provide a deep study of two approaches for generating differential updates --- namely, incremental and cumulative --- and compare their efficiency. We show that the incremental approach allows significantly reducing the amount of generated control information compared to the cumulative one, while providing the same level of information relevance. We develop an analytical model for the incremental approach and propose an algorithm which allows tuning its parameters, depending on the number of nodes in the network, their mobility, and wireless channel quality. Using the developed analytical model, we show that the  incremental approach is very useful for static dense network deployments and networks with low and medium mobility, since it allows us to significantly reduce the amount of control information compared to the classical full dump approach.
\end{abstract}


\vspace{-0.5cm}
\section{Introduction}
\label{sec:intro}
In many existing and emerging scenarios, it is highly necessary for the networking devices to have relevant information from their neighbors. For example, various routing protocols for mobile ad hoc networks (MANET) exchange information about links and their quality \cite{olsr-rfc} to distributively build a graph representing network topology. Outdated routing information may lead to loops and user data losses. Another example is dense deployment, where access points may reduce interference by coordinating transmissions in their networks with a sort of time division \cite{krasilov2013aa, tdma-krotov}, dynamic sensitivity control \cite{afaqui2016dynamic} or adaptive transmission power control \cite{kim2006improving} enabled by the novel IEEE 802.11ax amendment. Information dissemination is especially crucial for vehicular networks \cite{panichpapiboon2012review}.

Often, both user and control information share the same channel resources. Thus, efficient control information dissemination increases capacity for user traffic, in addition to improving scalability in terms of both the number of nodes which generate information updates and the rate of such updates. This makes the problem of control information dissemination crucial for narrow-band networks (see~\cite{fakhriev2013narrowband}) and networks with highly mobile devices. 

Typically, in ad hoc networks, control information is broadcast without being acknowledged. Because of error-prone nature of the wireless channel, broadcast messages can be lost. That is why in many protocols (e.g. OLSR~\cite{olsr-rfc}, RR-ALOHA~\cite{rr-aloha}), such broadcast messages contain \emph{full dumps} of information and are sent periodically, even if no changes occur in the network. The period of such messages is chosen as a trade-off between information relevance and channel time consumption. Despite the simplicity of the full dump approach, it is robust to packet losses since the neighboring nodes will recover lost information with the next successfully received message. Moreover, when a new node appears in the network, it receives all necessary information in the first received message. 
The cost for these benefits is a huge overhead (e.g. see~\cite{olsr-overhead}). 

Another approach referred to as \emph{group-based} one is proposed in the IEEE 802.11-2012 standard~\cite{802.11-2012} for the deterministic channel access protocol called Mesh coordination function Controlled Channel Access (MCCA, for details see~\cite{khorov2017will}). According to this approach, a node divides various pieces of information (information elements) into a relatively small number of groups and periodically sends information only about those groups that have been changed. Specifically, with MCCA protocol, a station periodically sends information about time intervals reserved for transmissions~\cite{ivanov2016modeling}. In~\cite{ISWCS2015,PIMRC2016} various group management algorithms are studied and compared in terms of the amount of control information, using the developed analytical models and simulations.  However, these papers do not take into account the relevance of information at neighboring nodes, since it is assumed that all control messages are transmitted reliably. 

The popular approach to reduce control information is to interleave full dump messages with short \emph{differential} updates, which contain only modified information. This approach can be implemented in two ways: with cumulative and incremental updates. The cumulative differential updates contain all information elements modified since the last full dump message. Such an approach is used by the DSDV~\cite{dsdv} routing protocol. In contrast, incremental differential updates contain only information elements modified since the previously transmitted message (a full dump or an incremental message). This approach is adopted by the OSPF-MDR~\cite{ospf-mdr} and the PSR~\cite{psr} routing protocols. Obviously, when all control messages are transmitted reliably, the cumulative approach produces more control information than the incremental one. However, a failure in differential update transmission for the incremental approach may lead to irrelevant information, while, for the cumulative approach, any successfully received control message makes the information relevant, except for the case when a full dump message has been lost. 

In this paper, we compare the two approaches and show that the incremental approach produce less control information compared to the cumulative one at the same information relevance level. We develop an analytical model of the incremental approach, which allows finding its optimal parameters in terms of the minimal amount of control information subject to some predefined probability of information relevance. Using the developed analytical model, we show that the  incremental approach is very useful for static dense network deployments and networks with low and medium mobility, since it allows significantly reduce the amount of control information compared to the classical full dump approach.

The rest of the paper is organized as follows. In Section~\ref{sec:scenario}, we specify the considered network scenario. Section~\ref{sec:prelim_sim} compares the cumulative and incremental approaches by means of simulation. In Section~\ref{sec:math_model}, we develop an analytical model of incremental approach which allows estimating the amount of generated control information and the probability of information relevance. Based on this model, we provide an algorithm for selecting the optimal parameters. In Section~\ref{sec:results}, we validate our model by simulation and evaluate the performance of the incremental approach and the developed algorithm using this model. Section~\ref{sec:conclusion} concludes the paper.      

\section{Scenario}
\label{sec:scenario}
   
In this paper, we consider the following scenario. Each node of a wireless network periodically broadcasts to its neighbors control messages of two types: \emph{full dump} and \emph{differential update} messages. We refer to the period of control message transmission as a \emph{slot}. We assume that a node transmits a control message at the beginning of each slot. To reduce the amount of sent control information, the node interleaves differential update messages and the full dump ones. Let $N$ be the transmission period for the full dump messages, i.e., a node transmits full dumps at the beginning of every $N$-th slot, while all other messages are differential updates.

The size of control messages and their loss probabilities depend on the rate of appearing new information elements (e.g., appearing of new scheduling information for channel access protocols or new links and metric updates for routing protocols) and their lifetimes. In this paper, we assume that the number of information elements appearing at a node in a time slot has a Poisson distribution with rate $\lambda$. The lifetime of each information element has an exponential distribution with mean $1/\mu$ slots. The size of each information element is constant and equals $V_0$ bits. Similar to our previous works~\cite{ISWCS2015,PIMRC2016}, we assume that the average lifetime of an information element is much longer than one slot, i.e. $1 / \mu \gg 1$. Also, we limit the number of information elements tracked by each node with threshold $R$.

Let us consider a node (referred to as node A) having $M$ neighbors. We assume that the wireless channel is error prone and the probability that neighbor $i$ cannot decode a control message from node A containing $s$ information elements equals $p_{err}^{(i)}(s)$. For definiteness, we estimate $p_{err}^{(i)}(s)$ as follows: 
\vspace{-0.2cm}
\begin{equation}
	\label{eq:p_err}
p_{err}^{(i)}(s)=1-(1-ber^{(i)})^{sV_0},
\end{equation}
where $ber^{(i)}$ is the Bit Error Rate (BER) at node $i$. However, any other dependency can be considered as well. 

We assume that the network topology is dynamic, i.e., the nodes are mobile. Hence, node A loses connections with its neighbors when they move far away and establishes new connections with the nodes appearing in its coverage area. In this paper, we assume that the duration of the connected phase for each neighbor (i.e. the time interval during which it is connected to node A) has the exponential distribution with mean $1/\gamma$.
   
When a neighbor fails to receive an update from node A, the information at this neighbor becomes irrelevant. To obtain up-to-date information, the neighbor needs to successfully receive a full dump message from node A in case of the incremental approach or any type of control message in case of the cumulative one. Let us define information relevance probability as the probability that at any arbitrarily selected time slot all \emph{connected} neighbors have relevant information generated by node A. To provide correct operation of a protocol which disseminates control information, we should guarantee that the relevance probability of that control information is higher than some threshold $p_{thresh}$.       

To increase the relevance probability, nodes use unsolicited retries scheme. It means that nodes broadcast messages several times in row. Let us denote $n_f$ and $n_d$ the number of transmission attempts for the full dump and differential update messages, respectively. These two variables together with period $N$ of full dump messages are considered as the parameters and can be tuned in order to minimize the amount of generated control information and to keep the relevance probability above the threshold $p_{thresh}$. We estimate the amount of control information as the average number of information elements which are sent at the beginning of a slot.
   
\section{Preliminary analysis}
\label{sec:prelim_sim}

In this section, we compare incremental and cumulative approaches for generating differential updates. For that, we use an event-driven custom simulation program, run experiments in the scenario described in Section~\ref{sec:scenario}, and average the obtained results over $20$ simulation runs. We consider transmission of control information via a single link (i.e. $M=1$). For higher values of $M$, we obtain similar results.  Unless otherwise stated, further we set: $R=1000$, $\gamma = 0.001$, $V_0 = 2$~bytes. BER on the considered link is set to $ber=6.6\cdot10^{-6}$, which means that the probability of incorrect reception of a message containing $R$ information elements equals $p_{err}(R)\approx10\%$.

\begin{figure}[!t]
	\vspace{-0.5cm}
	\centering
	\includegraphics[width=\linewidth]{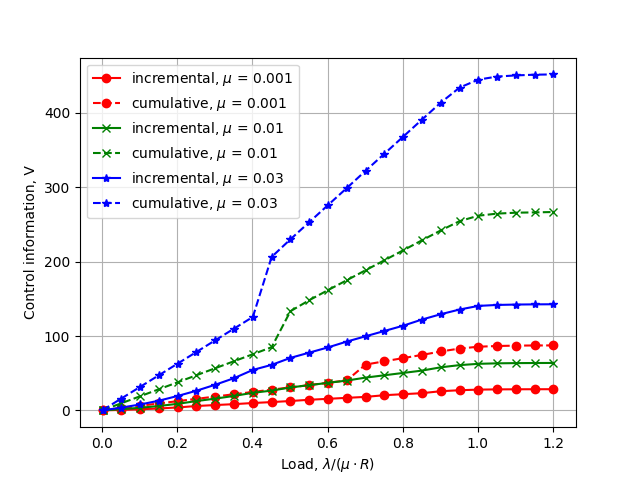}
	\caption{\label{fig:cmp} Comparison of incremental and cumulative approaches.}
	\vspace{-0.5cm}
\end{figure}

For each of two approaches, we find via the exhaustive search such triple $(N^{*}, n_d^{*}, n_f^{*})$ that minimizes the amount of control information subject to the relevance probability being higher than $p_{thresh}=0.95$. Fig.~\ref{fig:cmp} shows the average amount of control information for optimal triple (for each $\lambda$ and $\mu$) as a function of the load defined as $\frac{\lambda}{\mu R}$.  We can see that for the load higher than $1.0$, the amount of  control information does not change since the number of information elements tracked by node A is close to $R$ and cannot further increase.
Note that the curves corresponding to the cumulative approach experience a significant growth at the load of $0.4\:..\:0.5$. At this point, the cumulative approach --- which is more robust to packet losses by design --- needs to increase the number of retries for full dumps and/or differential updates  in order to satisfy the given requirement on the information relevance probability. Certainly, the incremental approach also increases the number of retries with the load. However this leads only to a slight increase of the amount of control information, since the size of differential update messages for the incremental approach is much lower than for that of the cumulative one. 

From the presented above results, we can conclude that the incremental approach outperforms the cumulative one: for all the considered loads and $\mu$ values, the incremental approach generates three times less control information. So, further in this paper, we focus on the incremental approach and develop an analytical model which allows selecting its optimal parameters (i.e. $N$, $n_d$ and $n_f$). 

\section{Analytical model}
\label{sec:math_model}

Let us develop an analytical model of the incremental approach which allows estimating the average amount of generated control information and the information relevance probability. Based on this model, we propose an algorithm to optimize its performance in terms of minimizing the amount of control information subject to a given requirement on information relevance.

\subsection{Estimation of the average amount of control information}
\label{subsec:ad_info}
   
Let us consider node A and estimate the average amount of the generated control information. For that we need to find stationary probabilities $\pi_{r}$ that the node has $r$ information elements. In our previous paper~\cite{ISWCS2015}, we have developed an analytical model that allows estimating the average amount of generated control information for the full dump approach (i.e. for the case when $N=1$). The model is based on a discrete time Markov chain with state $r$ and the time unit equal to the slot. For that chain, we have found the stationary probabilities $\pi_{r}$, which we further use in this paper. 
   
Following~\cite{ISWCS2015}, let us introduce the following notation: $d$ is the number of information elements deleted in the current slot;  $n$ is the number of new information elements added in the current slot. 

Since the lifetime of each information element has exponential distribution, the probability of deleting a particular information element during a slot equals $\tilde{p} = 1 - e^{-\mu}$. As lifetimes of different information elements are mutually independent random variables, the probability of deleting $d$ out of $r$ information elements equals   
\begin{equation}
   \label{eq:p_dr}
   p_{d|r} = C_{r}^{d} \tilde{p}^{d} (1 - \tilde{p})^{r-d}. 
\end{equation}
   
Using $\pi_{r}$ and~\eqref{eq:p_dr}, we can find the average numbers $\avg{r}$ and $\avg{d}$ of the total and deleted information elements, respectively:
\vspace{-0.5cm}
\begin{align*}
 &  \avg{r} = \sum\limits_{r=0}^{R} r  \pi_{r},  \\  
 &  \avg{d} = \sum\limits_{d=0}^{R} d  \sum\limits_{r=d}^{R} \pi_{r} \cdot p_{d|r}.
\end{align*}
   
At the steady state, the average number of information elements $\avg{n}$ added during a slot equals the average number of information elements $\avg{d}$ deleted during a slot. Since every differential update message contains information elements, which are added and deleted during the slot, the average size of a differential update message equals $2 \avg{d}$. Since a full dump message contains all information elements, its average size  equals $\avg{r}$. Taking into account that each full dump message and each differential update message are repeated $n_f$ and $n_d$ times, respectively, we can estimate the average amount of control information as follows:

\vspace{-0.3cm}   
\begin{equation}
   \label{eq:avg_V}
   \avg{V} = \frac{1}{N} n_f \avg{r} + 2\frac{N-1}{N} n_d  \avg{d}.
\end{equation}
   
\subsection{Information relevance probability}
\label{subsec:actuality_prob}
   
Now let us estimate the information relevance probability. First, we consider a pair of nodes (e.g. node A transmits control messages to node B) and estimate the probability $p_{rel}$ that the control information obtained at node B is relevant. After that, we generalize the result for the case when node A has several neighbors.  
 
Let us calculate the information relevance probability within the connected phase of node B (see Fig.~\ref{fig:active_cycle}). For that, we consider $N$ consecutive slots (the full dump cycle) and assume that at the beginning of the first slot, node A sends a full dump message.
We can consider different full dump cycles independently since only the full dump message sent at the beginning of each full dump cycle can recover information relevance at node B. Hence, we only need to estimate the relevance probability $\hat{p}_{rel}$ within one full dump cycle. 
      
From~\cite{ISWCS2015}, the conditional probability of adding $n$ new information elements for given $r$ and $d$ values equals:   
\begin{equation}
   \label{eq:p_nrd}
   p_{n|r, d} =
   \begin{cases}
   p_k(n), & n < R + d - r,\\
   1 - \sum\limits_{k = 0}^{n - 1} p_k(k), & 0 < n = R + d - r,\\
   1, & 0 = n = R + d - r,
   \end{cases}
\end{equation}
where $p_k(k) = \frac{\lambda^{k}}{k!} e^{-\lambda}$.
   
Using stationary probabilities $\pi_{r}$ and equations \eqref{eq:p_dr} and~\eqref{eq:p_nrd}, we can estimate loss probability  $p_{f}$ after $n_f$ retries of a full dump message and loss probability $p_{d}$ after $n_d$ retries of a differential update message as follows:

\vspace{-0.1cm}   
\begin{equation*}
   \label{eq:p_f}
   p_{f} = \left[ \sum\limits_{r = 0}^{R} \pi_{r}  \cdot p_{err}(r) \right]^{n_f},
\end{equation*}
\vspace{-0.2cm}
\begin{align*}
   \label{eq:p_d}
   p_{d} &= \left[ \sum\limits_{r = 0}^{R} \pi_{r} \sum\limits_{d = 0}^{r} p_{d|r} \sum\limits_{n = 0}^{R - (r - d)} p_{n|r,d} \cdot p_{err}(d+n)\right]^{n_d},
\end{align*}
where $p_{err}(r)$ is calculated according to equation \eqref{eq:p_err} (we omit index $i$ here).

If a full dump message is lost, the information at node B is irrelevant for the whole full dump cycle, i.e. $N$ slots. Otherwise, if the $i$-th differential update (out of $N-1$ differential updates) is the first lost control message within the full dump cycle, the information at node B is relevant for $i$ time slots and irrelevant for the other $N-i$ slots. Hence, the relevance probability within one full dump cycle equals $\hat{p}_{rel} = p_{f} \cdot 0 + (1 - p_{f}) \cdot \left[ \sum\limits_{i = 1}^{N-1} (1 - p_{d})^{i-1} p_{d} \cdot \frac{i}{N} + (1-p_{d})^{N-1} \cdot 1 \right]$. After summing up, we obtain:

\vspace{-0.2cm} 
\begin{equation}
   \label{eq:hat_p_act}
   \hat{p}_{rel} = \frac{(1 - p_{f})(1 - (1 - p_{d})^{N})}{N p_{d}}.
\end{equation}
   
According to the considered scenario, node B can be in two states, connected and disconnected. The average duration of connected phase equals $\frac{1}{\gamma}$. Since the duration of the full dump cycle $N$ should be small enough compared to the duration of connected phase to meet information relevance requirement $p_{rel} \geq p_{thresh}$ (note that $p_{thresh}$ is close to $1$), we can write $\frac{1}{\gamma} >> N$. Hence, the probability that node $B$ has relevant information subject to the condition that node $B$ is in connected state and node $A$ has already sent full dump message equals $\hat{p}_{rel}$ calculated according to equation~\eqref{eq:hat_p_act}. The startup period, i.e. the period between the time when node B connects to node A and the time of first full dump message transmitted by node A (see Fig.~\ref{fig:active_cycle}) has uniform distribution over the interval $[0, N]$ with the mean value $\frac{N}{2}$. Thus, the information relevance probability at node B --- when it is in the connected state --- can be calculated as follows:

\begin{figure}[!t]
	\vspace{-0.5cm}
   	\centering
   	\includegraphics[width=0.5\textwidth]{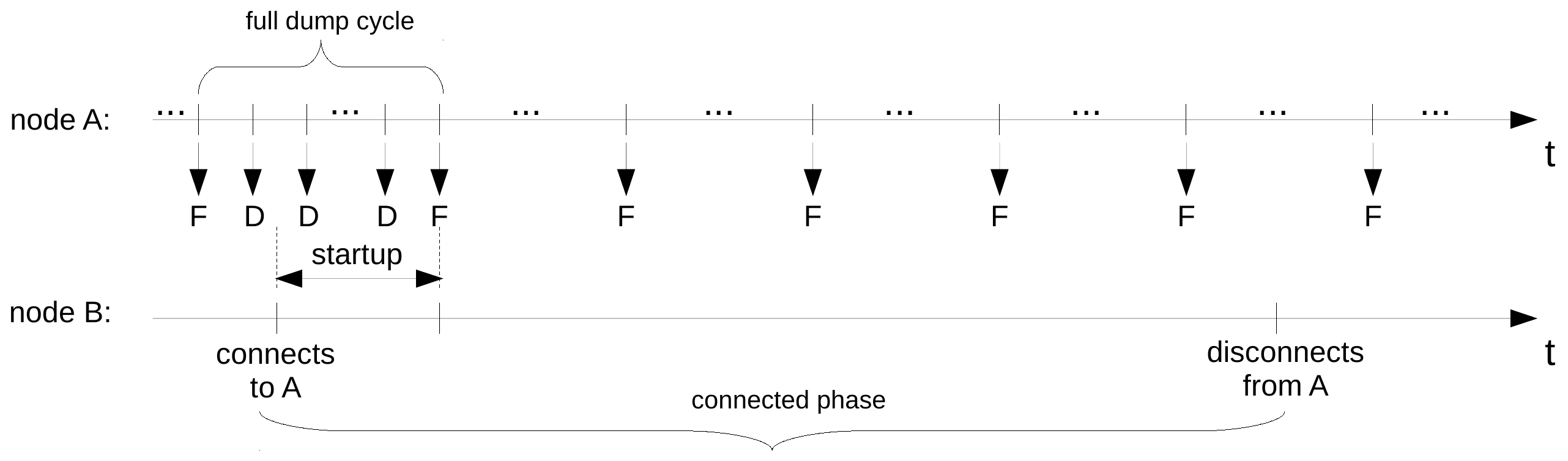}
   	\caption{\label{fig:active_cycle} Connected phase.}
   	\vspace{-0.5cm}
\end{figure} 

\begin{equation}
   \label{eq:p_act}
   p_{rel} = \hat{p}_{rel} \Big( \frac{1}{\gamma} - \frac{N}{2} \Big)/\frac{1}{\gamma}.
\end{equation}

Now let us consider the case when node A disseminates control information to $M$ neighbors. Assuming that all neighbors process messages and move independently, we can calculate the probability that the control information is relevant at \emph{all} neighbors as follows:
\begin{equation*}
	\label{eq:p_act_M}
	p_{rel}^{(all)} = \prod\limits_{i = 1}^{M} p_{rel}^{(i)},
\end{equation*} 
where $p_{rel}^{(i)}$ is calculated according to equation~\eqref{eq:p_act}.

\subsection{Asymptotic analysis}
\label{subsec:asymptotics}
   
Let us study the asymptotic case when $\lambda \rightarrow \infty$. In this case, the total number (and the average number $\avg{r}$) of information elements $r$ in every slot is constant and equals $R$. 
   
When in a particular slot $d$ information elements are deleted, the same number of information elements are added. Hence, for the loss probabilities $p_{f}$ and $p_{d}$ and the average number of deleted information elements per  slot $\avg{d}$ we have:
\vspace{-0.2cm}
\begin{equation*}
   \label{eq:p_f_as}
   p_{f} = [  p_{err}(R) ]^{n_f}, \;\; p_{d} = \left[ \sum\limits_{d = 0}^{R} p_{d|R} \cdot p_{err}(2d) \right]^{n_d},
\end{equation*}
\vspace{-0.2cm}
\begin{equation*}
   \label{eq:avg_d_as}
   \avg{d} = (1 - e^{-\mu}) R.
\end{equation*}

Therefore, the average amount of control information can be recalculated as follows:
\begin{equation*}
\label{eq:avg_V_as}
\avg{V} = \frac{1}{N} n_f R + 2\frac{N-1}{N} n_d  R (1 - e^{-\mu}).
\end{equation*}

Thus, for the considered asymptotic case, we obtain closed-form expressions which allow estimating  the average amount of control information and the relevance probability without need in solving the Markov chain and finding stationary probabilities $\pi_r$, which significantly reduces computational complexity. 
   
\subsection{Tuning parameters}
\label{subsec:params}
   
The performance of the incremental approach depends on its parameters (i.e. $N$, $n_f$ and $n_d$). 
These parameters should be selected in order to minimize the amount of control information $V$ subject to $p_{rel}^{(all)} \geq p_{thresh}$. The latter inequality imposes a restriction on $N$, since for each node the probability that information is irrelevant cannot be less than the ratio of the duration of startup period and the duration of connected phase. Hence, the following condition for $N$ should hold: $(1 - \frac{N/2}{1/\gamma})^{M} \geq p_{thresh}$. Thus, the maximal value $N_{max}$ for parameter $N$, for which the condition $p_{rel}^{(all)} \geq p_{thresh}$ holds, can be found as follows:
\begin{equation}
\label{eq:N_max}
N_{max} = \left\lfloor \frac{2 (1 - \sqrt[M]{p_{thresh}})}{\gamma} \right\rfloor.
\end{equation}
   
We propose the following algorithm for selecting the optimal parameters for the incremental approach:
   
\begin{enumerate}
   	\item Find the maximal value $N_{max}$ according to~\eqref{eq:N_max}.
   	\item Find triples $(N, n_f, n_d)$, for which condition $p_{rel}^{(all)} \geq p_{thresh}$ holds, e.g., using the exhaustive search ($N = \{1..N_{max}\}$, $n_{f}$ and $n_d$ within reasonable limits, e.g. $n_f, n_d = \{1..7\}$), 
   	\item Choose from the triples found at step 2 the triple $(N^{*}, n_{f}^{*}, n_{d}^{*})$  providing the minimal amount of control information $V$ according to~\eqref{eq:avg_V}.  
\end{enumerate}
   
In Section~\ref{sec:results}, we also consider triple $(\tilde{N}, \tilde{n}_f, \tilde{n}_d)$, which is found with the same algorithm for the case $\lambda \rightarrow \infty$, and show that this triple provides close to optimal results.   
   
\section{Numerical results}
\label{sec:results}

\subsection{Model validation}
\label{subsec:validation}

Let us estimate the accuracy of the analytical model developed in Section~\ref{sec:math_model}. For that, we vary $\mu$ and load, and compare the average amount of control information sent per slot obtained with the analytical model and simulations. We consider the optimal parameter values $(N^{*}, n_{f}^{*}, n_{d}^{*})$, which are chosen according to the algorithm in Section~\ref{sec:math_model}. We also consider the optimal parameter values $(\tilde{N}, \tilde{n}_{f}, \tilde{n}_{d})$ for the asymptotic case $\lambda \rightarrow \infty$. As in the preliminary analysis presented in Section~\ref{sec:prelim_sim}, we validate the analytical model in a single link scenario. We set threshold $p_{thresh} = 0.95$, mobility $\gamma = 0.001$ and BER corresponding to $p_{err}(R) = 10\%$. 

\begin{figure}[!t]
	\vspace{-0.5cm}
	\centering
	\includegraphics[width=0.95\linewidth]{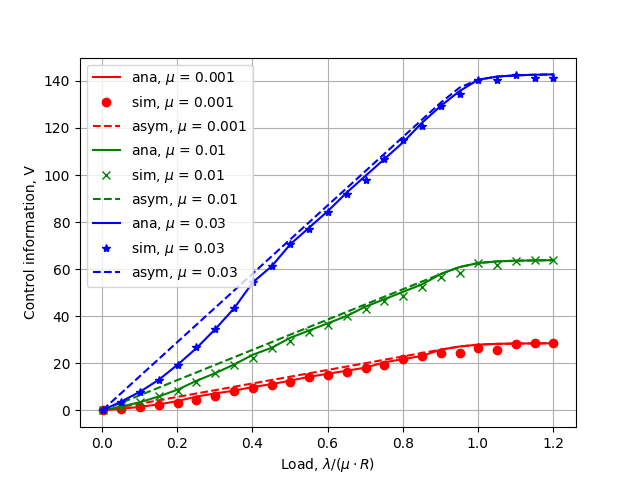}
	\caption{\label{fig:validation} Validation of the developed analytical model.}
	\vspace{-0.5cm}
\end{figure}

Fig.~\ref{fig:validation} shows that the curves obtained with analytical and simulation models almost coincide. Since the algorithm of selection $(N^{*}, n_{f}^{*}, n_{d}^{*})$ requires an accurate estimation for both information relevance probability $p_{rel}$ and the amount of control information $V$, we can conclude that the developed analytical model provides high accuracy. 

We can see that the amount of sent control information increases with the load and reaches its maximal value at high load (load $> 1.0$), where the estimations provided by the analytical model (solid line) and asymptotic analysis (dashed line) coincide. Note that parameter configuration $(\tilde{N}, \tilde{n}_f, \tilde{n}_d)$ provides results very close to optimal ones for medium and high load values. Specifically, for $\frac{\lambda}{\mu \cdot R} >= 0.5$ the amount of control information for both configurations is almost the same. Moreover, closed-form expressions obtained in Section~\ref{sec:math_model} allow tuning parameters online. Hence, further we investigate the incremental approach in the asymptotic case $\lambda \rightarrow \infty$.

\vspace{-0.15cm}
\subsection{Performance evaluation}
\label{subsec:perf_eval}

Now let us evaluate the performance of the incremental approach combined with the algorithm for tuning its parameters, based on the analytical model. The performance of the approach depends on the scenario parameters, such as mobility $\gamma$, number of neighboring nodes $M$, BER at each link $ber^{(i)}$ and information elements generation parameters $\lambda$ and $\mu$. Due to the lack of space, in all the experiments below we set $\mu = 0.01$. However, our experiments show that for other values of $\mu$ the results are quite similar.

\begin{figure}[!t]
	\vspace{-0.5cm}
	\begin{minipage}{\linewidth}
		\center{\includegraphics[width=0.95\linewidth]{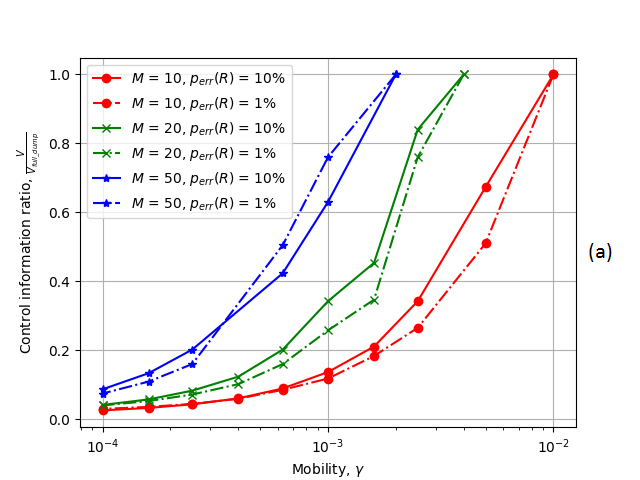}}\\
	\end{minipage}
	\vfill
	\vspace{-0.6cm}
	\begin{minipage}{\linewidth}
		\center{\includegraphics[width=0.95\linewidth]{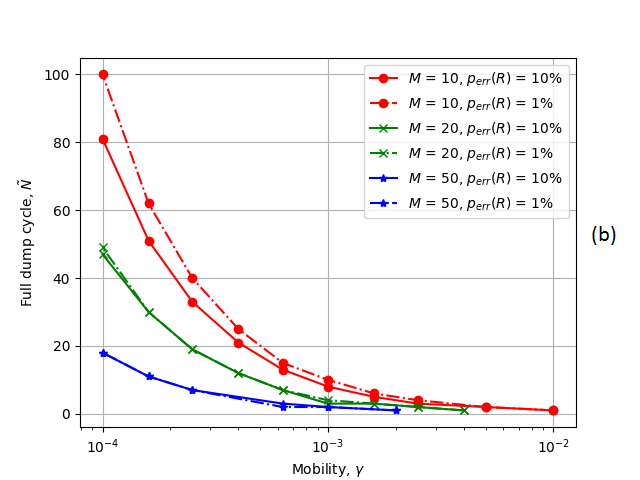}}\\
	\end{minipage}
	\caption{Performance of the incremental approach depending on the scenario parameters.}
	\label{fig:sensitivity}
	\vspace{-0.5cm}
\end{figure}

Fig.~\ref{fig:sensitivity} shows the dependencies of the amount of control information $V$ and optimal full dump cycle duration $\tilde{N}$ on mobility $\gamma$ for $M = \{10, 20, 50\}$. For all nodes, we set BER to the same value, which corresponds to $p_{err}(R) = \{1\%, 10\%\}$. Specifically, in Fig.~\ref{fig:sensitivity}(a) we consider the ratio of the amount of control information $V$ for the incremental approach with parameter configuration $(\tilde{N}, \tilde{n}_f, \tilde{n}_d)$ to the amount of control information $V_{full\_dump}$ for the full dump approach. Here, the full dump approach means that the node broadcasts full dump messages in every slot and does the minimal number of retries in order to satisfy requirement $p_{rel}^{(all)} \geq p_{thresh}$. Note that the full dump approach is the special case of the incremental one with the full dump cycle set to $1$. We can see that $V$ rises with $\gamma$ as higher mobility requires more frequent full dump message transmissions. Besides, we can see that $V$ also rises with BER since the node has to increase the number of retries $n_f$ and $n_d$ for the messages. At a particular level of mobility, the amount of control information generated by the node increases with the number of neighboring nodes, because it needs to maintain a given information relevance level at all neighbors. Fig.~\ref{fig:sensitivity}(b) shows that the full dump cycle duration $\tilde{N}$ decreases when mobility, BER and the number of stations increases.

It should be noted that $N_{max}$ calculated according to~\eqref{eq:N_max} can reach $0$, i.e., at some critical level of mobility, a node cannot further tune parameters to satisfy condition $p_{rel}^{(all)} \geq p_{thresh}$. Consequently, we can see that curves have no points in high mobility area $\gamma > \gamma_{critical}$. Specifically, the less the number of neighboring stations, the higher is the critical level $\gamma_{critical}$.

So, we can conclude that the incremental approach allows significantly reducing the amount of generated control information compared to the classical full dump approach, while providing a high level of information relevance for dense network deployments with low and medium mobility. 

\vspace{-0.2cm}
\section{Conclusion}
\label{sec:conclusion}
\vspace{-0.2cm}

In this paper, we have studied the problem of control information dissemination in error-prone wireless networks. We have compared two approaches (cumulative and incremental) for generating differential update messages  and have shown that the incremental approach generates significantly lower amount of control information compared to the cumulative one while providing the same level of information relevance. We have developed an analytical model which allows tuning parameters of the incremental approach in order to minimize the amount of generated control information and to meet a given requirement on information relevance. Numerical results have shown a high accuracy of the developed analytical model. In addition, for the asymptotic case we have provided closed-form expressions which allow finding suboptimal parameters with a low complexity algorithm. 

Using the developed analytical model, we have studied how the performance of the incremental approach depends on the number of nodes in the network, their mobility and wireless channel quality. The results have shown that at low and medium mobility the incremental approach significantly reduces the amount of generated control information compared to the classical full dump approach and at the same time provides a high level of information relevance. So, the incremental approach with the proposed algorithm for tuning its parameters is very useful for static dense network deployments and networks with low and medium mobility. 
 
In our further work, we are going to study the efficiency of the incremental approach combined with the proposed  algorithm for tuning its parameters using system level simulations (e.g. using NS-3). In particular, we are going to consider existing  channel access or routing protocols which require control information dissemination and more realistic mobility models. Also, we are going to compare the performance of different approaches for control information dissemination, including the group-based approach, in various scenarios.      

\vspace{-0.1cm}
\bibliographystyle{elsarticle-num}
\bibliography{main}

\end{document}